\documentclass[letterpaper, 10 pt, conference]{ieeeconf}
\IEEEoverridecommandlockouts                              

\usepackage[utf8]{inputenc}
\usepackage[tbtags]{amsmath}
\usepackage{nicefrac}
\usepackage{bbm}
\usepackage[subnum]{cases}
\usepackage{multicol}
\usepackage{amssymb, psfrag}
\usepackage{graphicx}
\usepackage{epstopdf}
\usepackage{amsmath,amssymb,amsfonts}

\usepackage{amsthm,balance}
\usepackage{enumerate, makecell}
\usepackage{graphicx, textcomp}
\usepackage{color,cite,hyperref,mathtools,booktabs}
\usepackage{cleveref}
\usepackage{algorithm,algpseudocode,balance}

\allowdisplaybreaks

\newtheorem{rmk}{Remark}
\newtheorem{assumption}{Assumption}

\DeclareMathOperator{\diag}{dg}

\newcommand \mcE{\mathcal{E}}

\newcommand \mcG{\mathcal{G}}

\newcommand \mcV{\mathcal{V}}

\newcommand \mcZ{\mathcal{Z}}

\newcommand{\R}{\mathbb{R}}

\allowdisplaybreaks

\begin{document}

\title{Robust Decentralized Secondary Control Scheme \\ for Inverter-based Power Networks}

\author{Siddharth Bhela, Abhishek Banerjee, Ulrich Muenz, and Joachim Bamberger
}
	
	\maketitle
	\thispagestyle{empty}
	\pagestyle{empty}

\begin{abstract}
    Inverter-dominated microgrids are quickly becoming a key building block of future power systems. They rely on centralized controllers that can provide reliability and resiliency in extreme events. Nonetheless, communication failures due to cyber-physical attacks or natural disasters can make autonomous operation of islanded microgrids challenging. This paper examines a unified decentralized secondary control scheme that is robust to inverter clock synchronization errors and can be seamlessly applied to grid-following or grid-forming control architectures. The proposed scheme overcomes the well-known stability problem that arises from parallel operation of local integral controllers. Theoretical guarantees for stability are provided along with criteria to appropriately tune the secondary control gains to achieve good frequency regulation performance while ensuring fair power sharing. The efficacy of our approach in eliminating the steady-state frequency deviation is demonstrated through simulations on a 5-bus microgrid with four grid-forming inverters.
\end{abstract}

\section{Introduction} \label{sec:introduction}
	
Microgrids (MG) consist of a group of interconnected distributed energy resources and loads that act as a single controllable entity. Since MGs can operate in grid-connected or islanded modes they are touted as the key building blocks of future power systems~\cite{MGs_importance,IBR_future}. Inverter-based MGs are attracting attention in industry and academia as they can improve reliability and ensure support for critical services even during extreme events \cite{MGM08}. Operation of MGs in islanded mode is considered challenging as the dynamics of the MG are no longer dominated by the main grid. In such scenarios, advanced control mechanisms are needed to maintain the delicate demand-supply balance~\cite{MG_overview}.

Hierarchical control schemes have been well-explored for operation of inverter-dominated microgrids \cite{MC10,heirarchical_control1}. These schemes are classified into three levels of control that serve different functions: $i)$ the primary control layer is the fastest and establishes power sharing; $ii)$ the secondary control layer is responsible for providing frequency regulation and eliminating steady-state frequency deviations introduced by the primary control; and $iii)$ the tertiary control layer is concerned with defining the the long-term set points based on economic dispatch~\cite{heirarchical_control2}.

The primary control layer is largely droop-based and relies purely on local measurements \cite{droopLV}, \cite{Divan93}, \cite{pesgm_2022}. However, the secondary and tertiary control layers typically depend on communication. While centralized control architectures for secondary control provide good performance they are neither scalable nor robust to cyber-physical attacks \cite{MC18}. For this reason, a variety of distributed and decentralized secondary control (DSC) schemes have been explored; see \cite{KH2020} and references therein. Despite their many benefits, communication-free control schemes in MGs can lead to poor performance and instability if the inverter digital processor clocks used to generate the time signals are not synchronized. The impact of clock synchronization on frequency regulation and power sharing has been briefly reviewed in literature \cite{MC16}, \cite{MC18}, \cite{PM18}, \cite{Dorfler16}. Nonetheless, there is no unified and robust DSC scheme that can be implemented in both Grid-forming (GFM) and Grid-following (GFL) inverters. Moreover, little thought is given on how to tune the secondary controller gains. In this paper, we address both these challenges.

Our contributions are as follows. First, in Section~\ref{sec.CFSCsolution} we provide a unified modeling framework for investigating DSC schemes for both GFM and GFL inverters. We also show that an adhoc DSC approach based on local integral controllers is not robust to clock synchronization errors. Moreover, a novel DSC scheme with damping is proposed that overcomes these challenges. Second, we provide stability guarantees for our proposed DSC scheme with damping in Section~\ref{sec.stability}. We show that both the secondary controller dynamics and the MG frequency reach a steady state. Further, based on the desired objectives conditions for appropriately choosing the secondary controller gains are also provided. Section~\ref{sec.NumericalTests} discusses simulation tests based on a 5-bus test case with four GFM inverters followed by conclusions and future research directions in Section~\ref{sec.Conclusion}.

\textbf{Notation:} Sets are denoted by calligraphic symbols. Given a real-valued sequence $\{x_{a,1}, \ldots, x_{a,N}\}$, $x_a$ is the $N \times 1$ vector obtained by stacking the entries $x_{a,i}$, and $\diag(x_a)$ is the corresponding diagonal matrix. The operator $(\cdot)^{\top}$ stands for transposition.


\section{Decentralized Secondary Control}
\label{sec.CFSCsolution}

An islanded microgrid having $N$ nodes can be modeled as a connected graph $\mcG = (\mcV, \mcE)$, whose nodes $\mcV :=
\{1,\ldots,N\}$ correspond to buses, and edges $\mcE$ to undirected lines. For simplicity of analysis we assume that there is an inverter at each bus $i \in \mcV$. Each inverter is equipped with the standard droop control \cite{Divan93} which is further augmented by the decentralized secondary control schemes presented in this section.

\begin{figure}[t]
	\centering
		\includegraphics[width=1.0\linewidth]{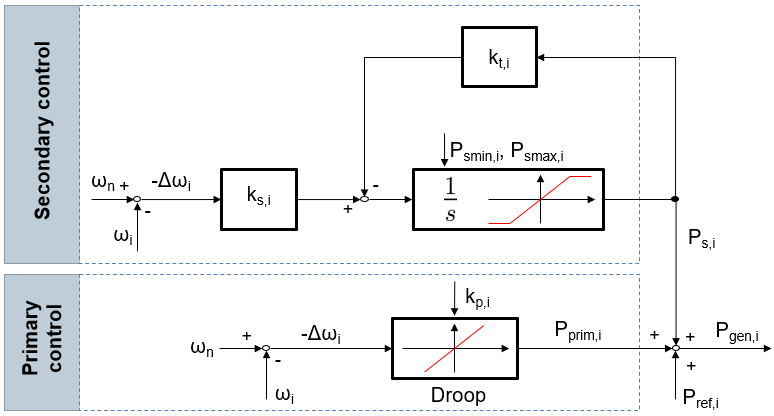}
	\caption{Control architecture for Grid-following (GFL) inverters.}
	\label{fig.powerControlledGenerator}
\end{figure}

\begin{figure}[t]
	\centering
		\includegraphics[width=1.0\linewidth]{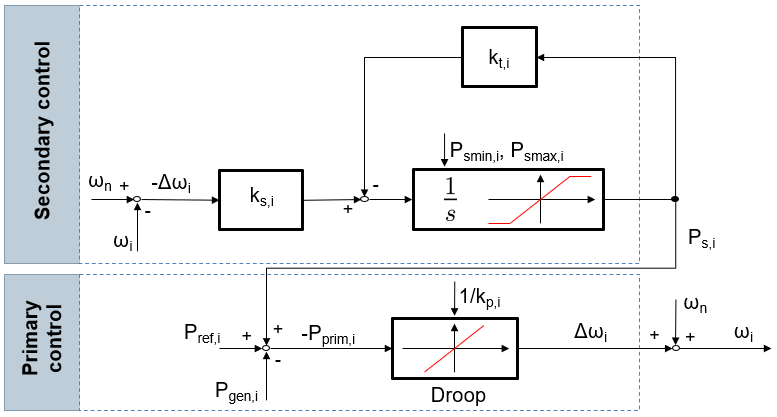}
	\caption{Control architecture for Grid-forming (GFM) inverters.}
	\label{fig.frequencyControlledGenerator}
\end{figure}

\subsection{Adhoc Decentralized Secondary Control}
\label{sec.adhocsolution}

The adhoc DSC solution consists of local integral controllers at all inverters. This corresponds to the control architectures shown in Figures \ref{fig.powerControlledGenerator} and \ref{fig.frequencyControlledGenerator} where the damping gain $k_{t,i}$ and any saturation are neglected. Under this simplification, let us consider the power balance \cite{kundur}
\begin{align}\label{eq:powerb}
  P_{gen,i} =& P_{ref,i} + P_{prim,i} + P_{s,i} \\
  =& P_{ref,i} + k_{p,i}(\omega_n - \omega_i) + P_{s,i} \nonumber \\
  =& P_{ref,i} - k_{p,i}\Delta \omega_i + P_{s,i} \nonumber
\end{align}
where $P_{gen,i}$, $P_{ref,i}$, $P_{prim,i}$, $P_{s,i}$, and $k_{p,i}$ are the real power output, reference power, primary reserve power, secondary reserve power, and primary droop gain of the inverter at bus $i$, respectively. \emph{Note that $k_{p,i}$ is the inverse of the standard droop gain}. The frequency at each bus $i \in \mcV$ is $\omega_i$ and the nominal frequency is denoted by $\omega_n$. Observe that \eqref{eq:powerb} without the corrective term $P_{s,i}$ is simply the primary droop equation \cite{kundur}, \cite{Divan93}. For notational convenience and to account for local loads let us define
\begin{subequations}
    \begin{align}
        P_i:=& P_{gen,i} + P_{load,i} \label{eq:Pi} \\ 
        P_{0,i}:=& P_{ref,i} + P_{load,i} \label{eq:P0}
    \end{align}
\end{subequations}
where $P_{load,i}$ is the total load at bus $i$. Heed that the derived model will describe the secondary control dynamics which are much slower than the primary control dynamics. Therefore, the frequencies measured across the grid can be assumed to be identical, i.e., $\omega_i \approx \omega^* \implies \Delta \omega_i=\Delta \omega=\omega^*-\omega_n$. By rearranging the terms in \eqref{eq:Pi} and \eqref{eq:P0} and substituting for $P_{gen,i}$ and $P_{ref,i}$ in \eqref{eq:powerb}, the power balance at each bus $i \in \mcV$ and collectively across the islanded microgrid can be expressed as
\begin{subequations}
   \begin{align}
        P_i =& P_{0,i}-k_{p,i}\Delta \omega + P_{s,i} = 0 \\
        1^{\top} P =& 1^{\top} P_0 - 1^{\top} k_{p} \Delta \omega + 1^{\top} P_{s}=0. \label{eq:powerb1}
    \end{align} 
\end{subequations}
Here $1$ is the $N \times 1$ vector of ones, and $P, P_0, k_p, P_s$ are the $N \times 1$ vectors obtained by respectively stacking the values $\{P_i\}, \{P_{0,i}\}, \{k_{p,i}\}, \{P_{s,i}\}$ at each bus. By rearranging the terms in \eqref{eq:powerb1}, the steady-state frequency can be inferred as
\begin{align}
  \Delta \omega =& \frac{1^{\top} P_0 + 1^{\top} P_{s}}{1^{\top} k_p}.
\end{align}
The adhoc solution can now be described by the following dynamical system
\begin{subequations} \label{eq.adhocModel}
    \begin{align} 
        \dot P_{s,i} =& - k_{s,i} \Delta \omega = - k_{s,i} \frac{1^{\top} P_0 + 1^{\top} P_s}{1^{\top} k_p} \\
        \dot P_s =& - \frac{1}{1^{\top} k_p} \left( k_s 1^{\top} \right) P_s - \frac{1^{\top} P_0 }{1^{\top} k_p} k_s
\end{align}
\end{subequations}

In practice, inverters operate with their own digital signal processors. The clocks used to generate the time signals differ from each other and without clock synchronization each inverter can have a different frequency offset $\gamma_i$ \cite{MC16}, \cite{MC18}. Accounting for this offset in the original model \eqref{eq.adhocModel} yields
\begin{subequations} \label{eq.adhocModelFrequencyError}
    \begin{align}
	   \dot P_{s,i} =& - k_{s,i} \left(\frac{1^{\top} P_0 + 1^{\top} P_s}{1^{\top} k_p} - \gamma_i \right) \label{eq.adhocModelFrequencyError1} \\
	   \dot P_s =& \underbrace{- \frac{1}{1^{\top} k_p} \left( k_s 1^{\top} \right)}_A P_s - \frac{1^{\top} P_0    }{1^{\top} k_p} k_s + \diag(k_s) \gamma \label{eq.adhocModelFrequencyError2}
\end{align}
\end{subequations}
The stability and robustness of this system on $\gamma$ depends on the system matrix $A$. Matrix $A$ is rank one by construction, i.e., $A$ has a zero eigenvalue with multiplicity $N-1$ and a non-zero eigenvalue. The eigenmode with nonzero eigenvalue has the eigenvector $k_s$ and the corresponding eigenvalue $- \frac{1^{\top} k_s}{1^{\top} k_p}$ since
\begin{align}
	A k_s =& - \frac{1}{1^{\top} k_p} \left( k_s 1^{\top} \right) k_s 
		= - \frac{1^{\top} k_s}{1^{\top} k_p} k_s .
\end{align}
Hence, the system has one stable eigenmode and all other eigenmodes are marginally stable \cite{Rugh96}. The second term in \eqref{eq.adhocModelFrequencyError2} is acting in the direction of the stable eigenmode of $A$. Therefore, sufficiently slow variations in $P_0$ are compensated by $P_s$. With respect to this input, the system \eqref{eq.adhocModelFrequencyError2} is stable and robust. The third term in \eqref{eq.adhocModelFrequencyError2} representing the frequency offset usually acts in different directions than the stable eigenmode of $A$. In fact, only homogeneous frequency offsets $\gamma = \overline \gamma 1, \overline\gamma\in\R$, are damped by the system matrix $A$. All non-homogeneous frequency offsets are continuously integrated by the parallel integral controllers. This shows that the adhoc solution is not robust to frequency offsets introduced due to clock synchronization errors \cite{Dorfler16}, \cite{MC16}.

\subsection{Decentralized Secondary Control with Damping}
\label{sec.slowDamping}

To counteract the unstable frequency offset dynamics reported in subsection \ref{sec.adhocsolution} we introduce an additional damping term $-k_{t,i} P_{s,i}$ as shown in Figures \ref{fig.powerControlledGenerator} and 
\ref{fig.frequencyControlledGenerator} to obtain
\begin{subequations} \label{eq.DSCD}
    \begin{align}
	   \dot P_{s,i} =& - k_{t,i} P_{s,i} - k_{s,i} \left(\frac{1^{\top} P_0 + 1^{\top} P_s}{1^{\top} k_p} -        \gamma_i \right) \\
	   \label{eq.slowDampingModel}
	   \dot P_s =& - \left( \diag(k_t) + \frac{1}{1^{\top} k_p} \left( k_s 1^{\top} \right) \right) P_s            \nonumber \\
		  & - \frac{1^{\top} P_0 }{1^{\top} k_p} k_s + \diag(k_s) \gamma.
\end{align}
\end{subequations}

The next section shows that this simple modification provides theoretical stability guarantees for an inverter dominated microgrid. Moreover, criteria are provided to appropriately tune the secondary control gains $(k_s, k_t)$ such that the steady-state frequency deviation is minimized. We also investigate how the ratio of aforementioned gains affects the secondary reserve unbalance and thereby the power sharing between inverters.
\section{Stability Analysis}
\label{sec.stability}
To study the stability of the proposed DSC scheme with damping we first investigate the steady-state of the dynamical system in \eqref{eq.DSCD} by setting $\dot P_{s,i} = 0$ to obtain
\begin{align}
  P_{s,i} =& \frac{k_{s,i}}{k_{t,i}}
      \left( - \frac{1^{\top} P_0 + 1^{\top} P_s}{1^{\top} k_p} 
        + \gamma_i \right).
\end{align}
The total secondary reserve power at steady-state can then be expressed as
\begin{subequations} \label{eq.totalSecondaryReserveSteadyState}
 \begin{align}       
  1^{\top} P_{s} &= - 1^{\top} k_{st}
    \frac{1^{\top} P_0 + 1^{\top} P_{s}}{1^{\top} k_p} 
      + k_{st}^{\top} \gamma \label{eq.totalSecondaryReserveSteadyState1}\\
    &= - \frac{1^{\top} k_{st}}
			{1^{\top} k_p + 1^{\top} k_{st} }1^{\top} P_0 + \frac{{1^{\top} k_p} k_{st}^{\top} \gamma}
			{1^{\top} k_p + 1^{\top} k_{st} } \label{eq.totalSecondaryReserveSteadyState2}
 \end{align}   
\end{subequations}
where $k_{st}$ refers to the vector with elements 
$\{\nicefrac{k_{s,i}}{k_{t,i}}\}$. The equality in \eqref{eq.totalSecondaryReserveSteadyState2} is derived by rearranging the terms in \eqref{eq.totalSecondaryReserveSteadyState1}. The steady-state frequency can now be computed as
\begin{subequations}
\begin{align}  
	\Delta \omega =& \frac{1^{\top} P_0 + 1^{\top} P_{s}}{1^{\top} k_p} \label{eq.frequencySteadyState0} \\
  =& \frac{1^{\top} P_0}{1^{\top} k_p} 
    - \frac{1^{\top} k_{st}}
			{1^{\top} k_p + 1^{\top} k_{st} } \frac{1^{\top} P_0}{1^{\top} k_p} + \frac{k_{st}^{\top} \gamma}
			{1^{\top} k_p + 1^{\top} k_{st} } \\
	\label{eq.frequencySteadyState}
  =& \frac{1^{\top} P_0}
			{1^{\top} k_p + 1^{\top} k_{st} }
		+ \frac{k_{st}^{\top} \gamma}
			{1^{\top} k_p + 1^{\top} k_{st} }
\end{align}
\end{subequations}
and the steady-state secondary reserve power as
\begin{subequations}
\begin{align}
  P_{s,i} =& \frac{k_{s,i}}{k_{t,i}}
      \left( - \frac{1^{\top} P_0}{1^{\top} k_p} - \frac{1^{\top} P_s}{1^{\top} k_p} 
        + \gamma_i \right) \label{eq.integratorSteadyState0}\\
   =& \frac{k_{s,i}}{k_{t,i}}
      \left( - \frac{1^{\top} P_0}{1^{\top} k_p} 
			+ \frac{1^{\top} k_{st}}
			{1^{\top} k_p + 1^{\top} k_{st} } \frac{1^{\top} P_0}{1^{\top} k_p} \right.\nonumber \\ 
	&\quad \left. - \frac{k_{st}^{\top} \gamma}
			{1^{\top} k_p + 1^{\top} k_{st} } + \gamma_i \right) \\
	\label{eq.integratorSteadyState}
   =& \frac{k_{s,i}}{k_{t,i}}
      \left( - \frac{1^{\top} P_0}
			{1^{\top} k_p + 1^{\top} k_{st} } - \frac{k_{st}^{\top} \gamma}
			{1^{\top} k_p + 1^{\top} k_{st} } + \gamma_i \right)
\end{align}
\end{subequations}
The equalities in \eqref{eq.frequencySteadyState} and \eqref{eq.integratorSteadyState} are obtained by substituting the expression for $1^{\top} P_s$ from \eqref{eq.totalSecondaryReserveSteadyState2} in \eqref{eq.frequencySteadyState0} and \eqref{eq.integratorSteadyState0}, correspondingly. We will revisit these equations later on. The subsequent analysis relies on the ensuing mild assumption that can removed with a more rigorous analysis.
\begin{assumption} \label{as1}
All damping gains are identical, i.e., $k_{t,i} = \bar k_t$ for all $i \in \mcV$.
\end{assumption}
To show that the damping term $\bar k_t$ actually damps the frequency offsets $\gamma_i$ we first transform the dynamic system \eqref{eq.DSCD}. We then separate the state space into a subspace that acts in the direction $k_s$ and an orthogonal subspace. Finally, we separate the dynamics in these two subspaces and show that the dynamics in both subspaces are stable.
Using the transformation
\begin{align} \label{eq.transformation}
	\tilde P_{s,i} =& \frac{P_{s,i}}{\sqrt{k_{s,i}}}
\end{align}
and the notation $\tilde k_{s,i}:= \sqrt{k_{s,i}}$ for the dynamic system in \eqref{eq.DSCD} yields
\begin{align} \label{eq.Pstilde0}
	\dot{\tilde P}_s =& \diag(\tilde k_s) \gamma - \diag(k_t) \tilde P_s
		\underbrace{- \frac{1}{1^{\top} k_p} \left( \tilde k_s \tilde k_s^{\top} \right)}_{\tilde A} \tilde P_s 
		- \frac{1^{\top} P_0 }{1^{\top} k_p} \tilde k_s.
\end{align}
As shown before in \eqref{eq.adhocModelFrequencyError2}, the matrix $\tilde A$ is rank one by construction and has a single 
non-zero eigenvalue
\begin{align}
	\tilde A \tilde k_s =& -\frac{1^{\top} k_s}{1^{\top} k_p} \tilde k_s.
\end{align}
Nevertheless, matrix $\tilde A$ is symmetric after this transformation and therefore all eigenvectors are orthogonal to each other \cite{Rugh96}. Especially, the eigenvectors corresponding to the zero eigenvalues are all orthogonal to $\tilde k_s$, that is
\begin{align}
	\tilde A v =& 0 \qquad \forall v \in \mcZ
\end{align}
and $\mcZ:=\{v\in\R^N : \tilde k_s^{\top} v = 0\}$. Let us now separate the state space of $\tilde P_s$ and the frequency offset $\gamma$ as follows
\begin{subequations}
\label{eq.separation}
\begin{align}
	\tilde P_s =& \alpha \tilde k_s + \hat P_s \label{eq.separation1}\\
	\gamma =& \bar \gamma 1 + \hat \gamma. \label{eq.separation2}
\end{align}
\end{subequations}
Here $\alpha$ is a scalar function, $\hat P_s$ is orthogonal to $\tilde k_s$, i.e., $\tilde k_s^{\top} \hat P_s = 0$ holds and $\hat \gamma$ 
is orthogonal to $k_s$, i.e., it satisfies $k_s^{\top} \hat\gamma = 0$. Note that the 
basis of the separation of $\gamma$ is not orthogonal and therefore we may have 
$\bar \gamma \ne 0$ even when $1^{\top} \gamma = 0$ because $1^{\top} \hat \gamma \ne 0$.
Pre-multiplying \eqref{eq.separation2} with $k_s^\top$ yields
\begin{subequations}
   \begin{align}
	k_s^{\top} \gamma =& k_s^{\top} 1 \bar \gamma \\
	\label{eq.barGamma}
	\implies \bar \gamma =& \frac{k_s^{\top} \gamma}{k_s^{\top} 1} \\
 \implies \hat \gamma =& \gamma - \frac{k_s^{\top} \gamma}{k_s^{\top} 1} 1.
\end{align} 
\end{subequations}
Therefore, $\bar \gamma$ is simply a weighted average of $\gamma$ with 
weights $\frac{k_{s,i}}{1^{\top} k_s}$. With the separation \eqref{eq.separation} and under assumption \ref{as1}
\begin{subequations} \label{eq.Pstilde}
\begin{align}
	\dot{\tilde P}_s =& \dot{\alpha} \tilde k_s + \dot{\hat P}_s \\
	=& - \bar k_t \left( \alpha \tilde k_s + \hat P_s \right)
		- \frac{1^{\top} k_s}{1^{\top} k_p} \alpha \tilde k_s 
		- \frac{1^{\top} P_0 }{1^{\top} k_p} \tilde k_s \nonumber \\
  & + \bar \gamma \tilde k_s + \diag(\tilde k_s) \hat \gamma \label{eq.Pstilde1}\\
	=& \left(- \bar k_t \alpha - \frac{1^{\top} k_s}{1^{\top} k_p} \alpha 
		- \frac{1^{\top} P_0 }{1^{\top} k_p} + \bar \gamma \right) \tilde k_s \nonumber \\
		& - \bar k_t \hat P_s + \diag(\tilde k_s) \hat \gamma. \label{eq.Pstilde3}
\end{align} 
\end{subequations}
The equality in \eqref{eq.Pstilde1} is obtained by substituting \eqref{eq.separation1} and \eqref{eq.separation2} in \eqref{eq.Pstilde0}. Note that now the first term in \eqref{eq.Pstilde3} is heading in the direction $\tilde k_s$ whereas the last two terms in the summand are orthogonal to $\tilde k_s$ because
$\tilde k_s^{\top} \hat P_s = 0$ and $\tilde k_s^{\top} \diag(\tilde k_s) \hat \gamma = k_s^{\top} \hat \gamma = 0$. Hence, we may separate the system of dynamic equations in \eqref{eq.Pstilde3} as follows
\begin{subequations}
\begin{align}
	\label{eq.secondaryReserveDynamics}
	\dot \alpha =& - \left( \bar k_t + \frac{1^{\top} k_s}{1^{\top} k_p} \right) \alpha 
		- \frac{1^{\top} P_0 }{1^{\top} k_p} + \bar \gamma \\
	\label{eq.unbalanceDynamics}
	\dot{\hat P}_s =& - \bar k_t \hat P_s + \diag(\tilde k_s) \hat \gamma .
\end{align} 
\end{subequations}
Notice that both dynamics are exponentially stable. As a last step, 
we recover our previously derived steady-states in 
\eqref{eq.frequencySteadyState} and \eqref{eq.integratorSteadyState}. Recall from the transformation in \eqref{eq.transformation} and the separation of state space in \eqref{eq.separation1} that 
\begin{subequations} \label{eq.Psnew}
\begin{align}
	P_s =& \diag(\tilde k_s) \tilde P_s \\
	=& \diag(\tilde k_s) \tilde k_s \alpha + \diag(\tilde k_s) \hat P_s \\
	=& k_s \alpha + \diag(\tilde k_s) \hat P_s.
\end{align} 
\end{subequations}
The steady-states of $\alpha$ and $\hat P_s$ can be derived from \eqref{eq.secondaryReserveDynamics} and \eqref{eq.unbalanceDynamics} as follows
\begin{subequations}
\begin{align}
	\alpha =& - \left(\bar k_t + \frac{1^{\top} k_s}{1^{\top} k_p} \right)^{-1} 
		\left(\frac{1^{\top} P_0}{1^{\top} k_p} - \bar \gamma \right) \\
	\label{eq.hatPsSteadyState}
	\hat P_s =& \frac{1}{\bar k_t} \diag(\tilde k_s) \hat \gamma.
\end{align}   
\end{subequations}
We investigate first the total secondary reserve power
\begin{align} \label{eq.1TPs0}
	1^{\top} P_s =& 1^{\top} k_s \alpha + \tilde k_s^{\top} \hat P_s = 1^{\top} k_s \alpha.
\end{align}
where the equality is obtained by substituting for $P_s$ from \eqref{eq.Psnew}. Since $\tilde k_s^{\top} \hat P_s = 0$ by construction of $\hat P_s$ we can simplify the previous equation as follows
\begin{subequations}
\begin{align}
	1^{\top} P_s = 1^{\top} k_s \alpha =& - \left(\frac{\bar k_t}{1^{\top} k_s} + \frac{1}{1^{\top} k_p} \right)^{-1} 
		\left(\frac{1^{\top} P_0}{1^{\top} k_p} - \bar \gamma \right) \\
	=& - \left(\frac{\bar k_t 1^{\top} k_p + 1^{\top} k_s}{1^{\top} k_s 1^{\top} k_p} \right)^{-1} 
		\left(\frac{1^{\top} P_0}{1^{\top} k_p} - \bar \gamma \right) \\
	=& - \frac{1^{\top} k_s 1^{\top} k_p}{\bar k_t 1^{\top} k_p + 1^{\top} k_s}
		\left(\frac{1^{\top} P_0}{1^{\top} k_p} - \bar \gamma \right) \label{eq.TSCd}
\end{align}    
\end{subequations}
Using the relation in \eqref{eq.barGamma} it is not hard to show that \eqref{eq.totalSecondaryReserveSteadyState} is equal to \eqref{eq.TSCd} under Assumption \ref{as1}.
We naturally assume that $\frac{1^{\top} P_0}{1^{\top} k_p} \gg \bar \gamma$ because 
$\frac{1^{\top} P_0}{1^{\top} k_p}$ describes the frequency deviation after the reaction
of the primary control which is certainly much larger than the weighted frequency offset $\bar \gamma$. Thus, we obtain
\begin{subequations}
\begin{align}
	1^{\top} P_s &\approx - \frac{1^{\top} k_s 1^{\top} k_p}{\bar k_t 1^{\top} k_p + 1^{\top} k_s}
		\frac{1^{\top} P_0}{1^{\top} k_p} \\
	=& - \frac{1^{\top} k_s}{\bar k_t 1^{\top} k_p + 1^{\top} k_s}
		1^{\top} P_0 . \label{eq.1TPs}
\end{align}    
\end{subequations}
Without damping ($\bar k_t = 0$) we recover the original solution $1^{\top} P_s = - 1^{\top} P_0$, i.e., eventually the secondary reserve compensates the power imbalance in the microgrid. If we include damping, the solution should be close to the original solution. This requires
that 
$\frac{1^{\top} k_s}{\bar k_t 1^{\top} k_p + 1^{\top} k_s}$ 
shall be close to one, which is achieved if 
\begin{align}
	\label{eq.tertiaryControlGainCondition}
	\bar k_t \ll \frac{1^{\top} k_s}{1^{\top} k_p} .
\end{align}
Given the expression in \eqref{eq.1TPs} the steady-state frequency deviation can now be determined as
\begin{subequations}
\begin{align}
	\Delta \omega =& \frac{1^{\top} P_0 + 1^{\top} P_s}{1^{\top} k_p} \\
	&\approx \frac{1}{1^{\top} k_p} \left( 1 - \frac{1^{\top} k_s}{\bar k_t 1^{\top} k_p + 1^{\top} k_s} \right) 1^{\top} P_0 \\
	=& \frac{\bar k_t {1^{\top} P_0}}{\bar k_t 1^{\top} k_p + 1^{\top} k_s}
\end{align}   
\end{subequations}
which corresponds to \eqref{eq.frequencySteadyState} under Assumption~\ref{as1}.
Note that \eqref{eq.tertiaryControlGainCondition} implies that the steady-state 
frequency deviation is close to zero, i.e. $\Delta \omega \approx 0$.

Finally, we compute the secondary reserve unbalance amongst the inverters in the microgrid. Recall from \eqref{eq.Psnew} that
\begin{subequations}
\begin{align}
	P_s =& k_s \alpha + \diag(\tilde k_s) \hat P_s \\
        =& k_s \alpha + \frac{1}{\bar k_t} \diag(k_s) \hat \gamma \label{eq.Psnew1}
\end{align}   
\end{subequations}
The desired steady-state is $P_s = k_s \alpha$ because this implies 
that the secondary control reserve is split up as specified by the 
secondary control gains $k_s$. Hence, the last term in \eqref{eq.Psnew1} should be as 
small as possible to achieve fair power sharing. 

\begin{rmk} \label{rmk1}
Notice from \eqref{eq.tertiaryControlGainCondition} and \eqref{eq.Psnew1} that there is an trade-off between the gains $(k_s, k_t)$. Larger ratios $\{k_{s,i}/k_{t,i}\}$ minimize the frequency deviation but compromise on power sharing. Similarly, smaller ratios $\{k_{s,i}/k_{t,i}\}$ improve power sharing, but provide poorer performance for frequency regulation. The gains can be tuned to achieve good frequency regulation performance while ensuring fair power sharing.
\end{rmk}
Finally, note that despite not considering a deadband in our secondary control architecture we were able to show that the dynamic system is stable and reaches a steady-state. 

\section{Simulation Results} \label{sec.NumericalTests}

\begin{figure}[t]
\centering
\includegraphics[scale=0.6,trim={0.1cm 0.1cm 0.1cm 0.1cm},clip]{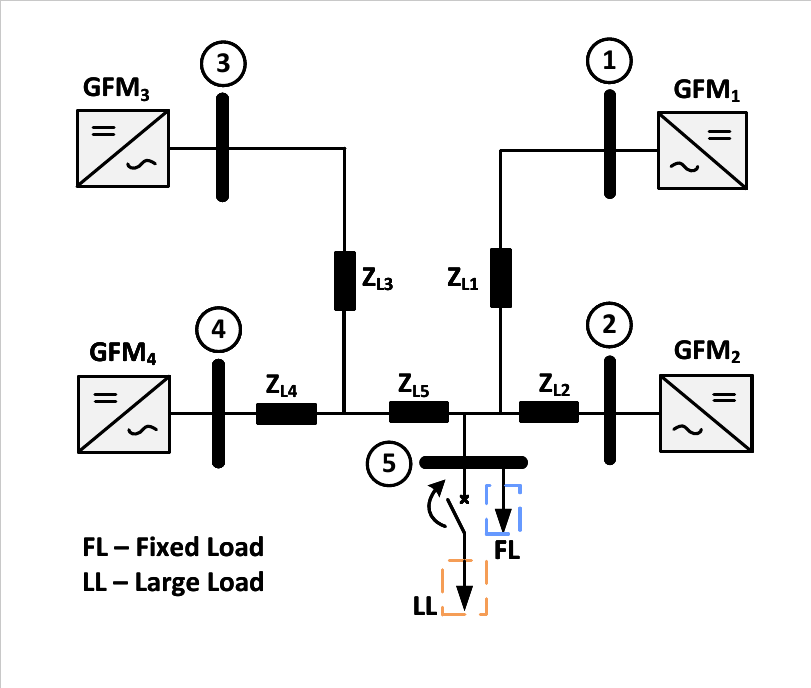}
\vspace{-15pt}
\caption{Single line diagram of the 4-inverter simulation testbed}
\label{fig:testbed}
\end{figure}

The simulation test system comprised of a microgrid (MG) with four GFM-inverters, a fixed load, and a variable large load. Lines connecting the inverters are modelled with impedances $Z_{Li}$. The single line diagram of the MG setup is shown in Figure~\ref{fig:testbed} and nominal values of the modeled MG components are provided in Table~\ref{table_1}. An averaged model of GFM inverters was used \cite{GFM_models}, and each inverter had the same LCL filter at its output stage with $(L_{f}, C_{f})$ defined in Table~\ref{table_1}. The inverters were programmed to operate in droop control mode with inner current and outer voltage control loops \cite{GFM_models}. 
This droop-based GFM control architecture was augmented with our DSC scheme.

\begin{figure}[h!]
\centering
\includegraphics[width=1.0\linewidth]{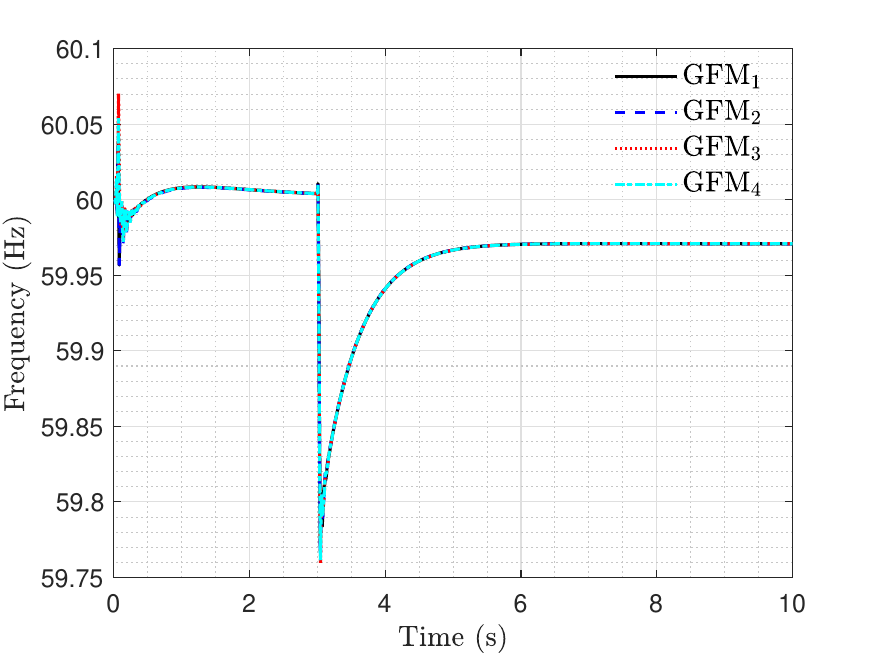}
\vspace{-10pt}
\caption{Frequency at the output of all GFMs for $k_{s}=1000$ and $k_{t}=0.25$}
\label{fig:all_freqs}
\end{figure}

\begin{table}[h!]
		\centering
		\caption{Nominal Values of the MG components}
		\begin{tabular}{ccc}
			\toprule
			Symbol & Description & Nominal value \\
			\cmidrule{1-3}
			$\omega_n$ &   Nominal Frequency  & $377$ $rad/s$\\
			$Z_{L1},Z_{L3}$ &  Line Impedance  & $1.6965+j0.9425~\Omega$     \\
		    $Z_{L2},Z_{L4},Z_{L5}$ &  Line Impedance  & $0.8482+j0.4712~\Omega$\\
            $L_{f}$ & All Inverter Filter L & $1.125$ $mH$\\
            $C_{f}$ & All Inverter Filter C & $11.5$ $\mu$ F\\
			$FL$ &  Fixed Load Power & $5$ $kW$   \\
			$LL$ &  Large Load Power & $2.5$ $kW$   \\
		    $P_{ref,i}$ &  Reference Power  & $2$ $kW$    \\
			$1/k_p$ &  Inverse Droop Gain & $1.5\%$   \\
                $T_{s}$ & Simulation Time Step & $20\mu s$ \\
			\bottomrule
		\end{tabular}
		\vspace{-0.5cm}
		\label{table_1}
\end{table}

\begin{figure}[h!]
\centering
\includegraphics[width=1.0\linewidth]{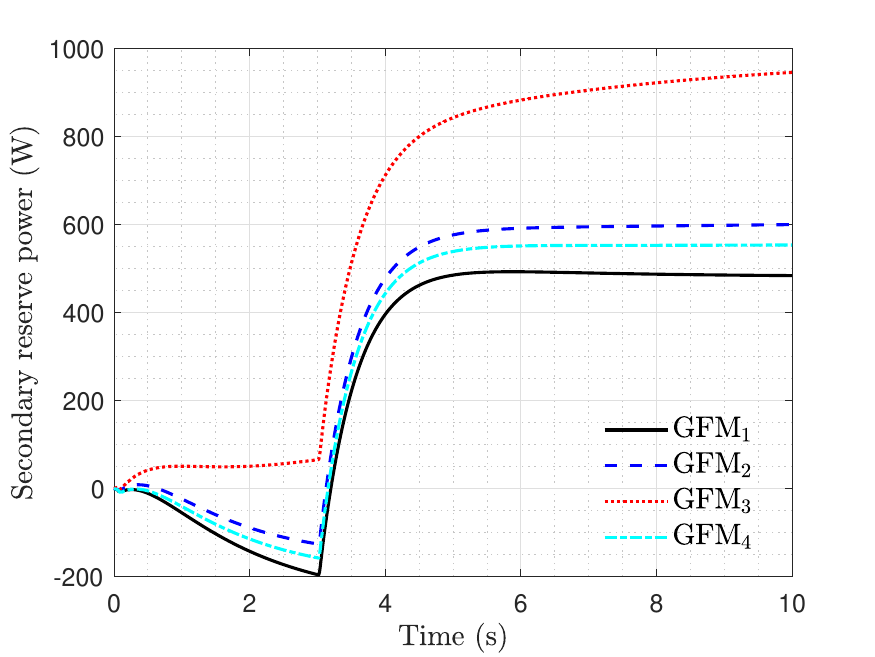}
\vspace{-10pt}
\caption{Secondary reserve power ($P_{s,i}$) for $k_{s}=1000$ and $k_{t}=0.25$}
\label{fig:all_Ps}
\end{figure}

\begin{figure}[h!]
\centering
\includegraphics[width=1.0\linewidth]{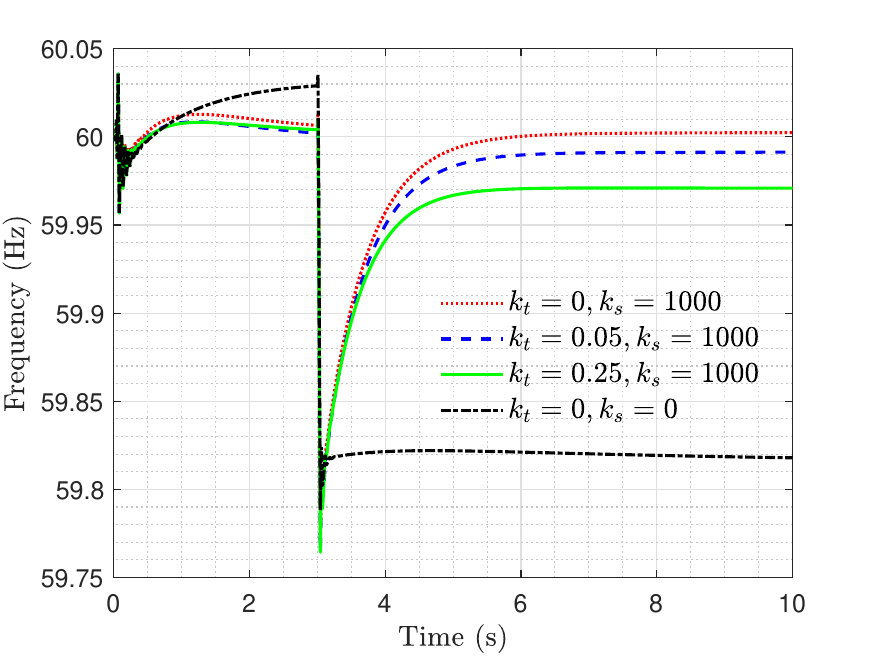}
\vspace{-10pt}
\caption{Frequency at $\textrm{GFM}_1$ for different values of $k_{t}$}
\label{fig:Kt_sweep_freqs}
\end{figure}

Performance of the proposed DSC scheme was tested in our simulation test system using MATLAB/Simulink \cite{MATLAB2020b}. The nominal voltage of the the system was $480$ $V$. For all test scenarios, the inverters were black started with a fixed load (FL) and a large load (LL) step was executed at $t=3$~seconds. Each GFM inverter was programmed to have a different frequency offset between $\pm 15$ $mHz$. Unless noted otherwise, the same gain values were used for all GFMs. For the first test scenario we chose the gain parameters ($k_s,k_t$) such that they satisfied criteria \eqref{eq.tertiaryControlGainCondition}. Figures \ref{fig:all_freqs} and \ref{fig:all_Ps} show that both the frequency and the secondary reserve power at all GFMs reached a steady-state. 

\begin{figure}[h!]
\centering
\includegraphics[width=1.0\linewidth]{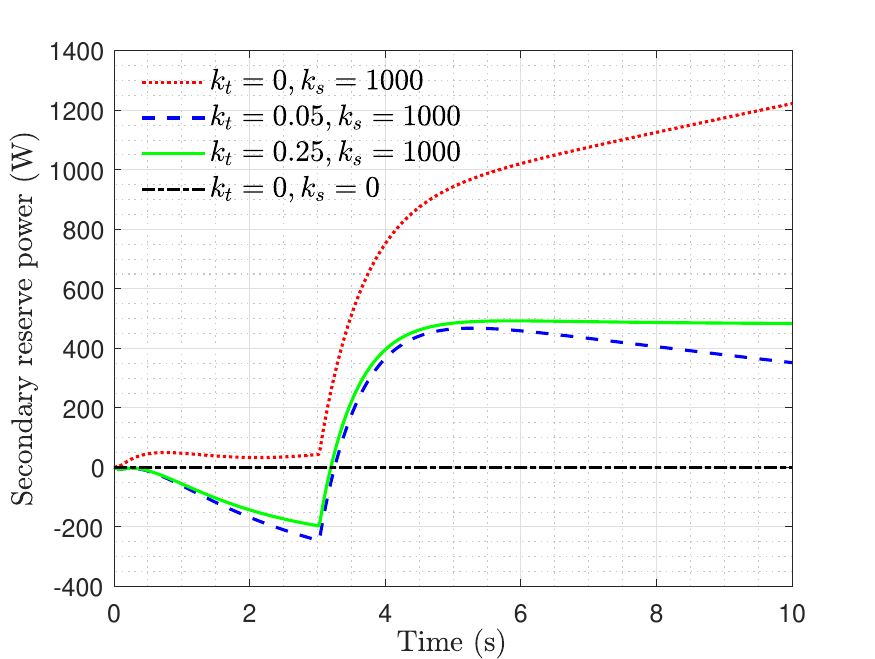}
\vspace{-10pt}
\caption{Secondary reserve power at $\textrm{GFM}_{1}$ for different values of $k_{t}$}
\label{fig:Kt_sweep_Ps}
\end{figure}

For the second test scenario we varied the gains $k_t$ to observe the response of the MG test system; see Figures \ref{fig:Kt_sweep_freqs} and \ref{fig:Kt_sweep_Ps}. Observe that for all positive pair of values $(k_s, k_t)$ the frequency and secondary reserved power reached a steady-state. For $k_t=0$ the secondary controller was continuously integrating the frequency offset, hence, the secondary reserve power was never able to reach a steady-state. For small values of $k_t \approx 0$ that satisfy \eqref{eq.tertiaryControlGainCondition} the steady-state frequency had smaller deviations from the nominal, however, this improvement in the frequency response came at the cost of increased settling time for the secondary reserve power dynamics. For the special case where no secondary control is involved, i.e., $k_s=0$ and $k_t=0$, the frequency settled at a much lower value.

\begin{figure}[h!]
\centering
\includegraphics[width=1.0\linewidth]{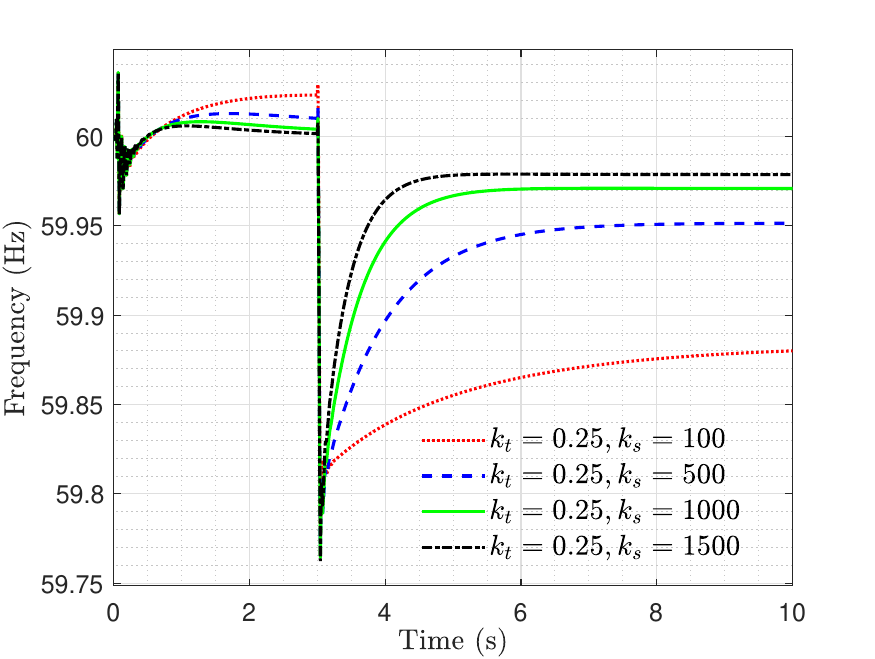}
\vspace{-10pt}
\caption{Frequency at $\textrm{GFM}_{1}$ for different values of $k_{s}$}
\label{fig:Ks_sweep_freqs}
\end{figure}

\begin{figure}[h!]
\centering
\includegraphics[width=1.0\linewidth]{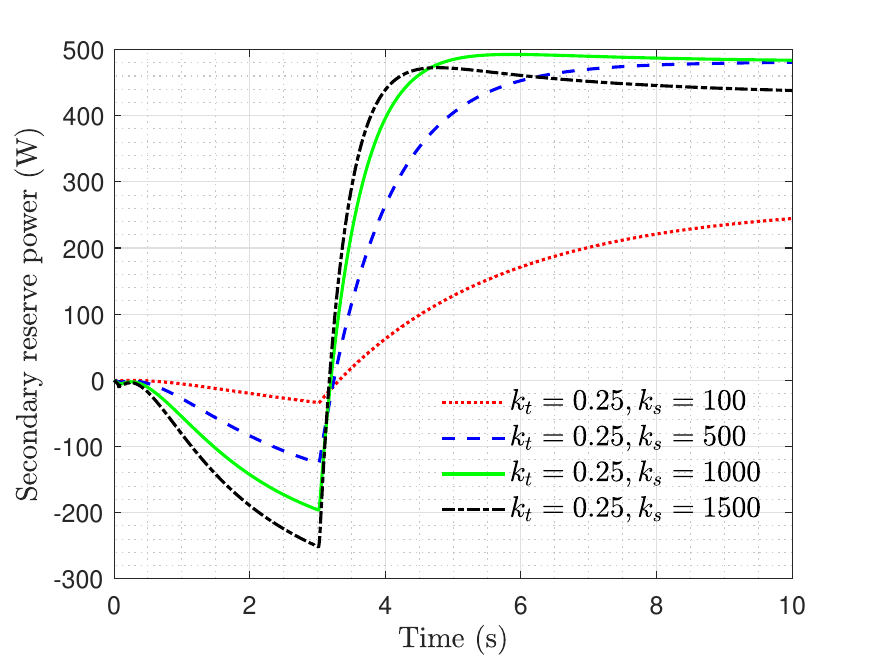}
\vspace{-10pt}
\caption{Secondary reserve power at $\textrm{GFM}_{1}$ for different values of $k_{s}$}
\label{fig:Ks_sweep_Ps}
\end{figure}

For the third test scenario we varied the gains $k_s$ to observe the response of the MG test system; see Figures \ref{fig:Ks_sweep_freqs} and \ref{fig:Ks_sweep_Ps}. Again, observe that for all pair of positive values $(k_s, k_t)$ both the frequency and the secondary reserve power reached a steady-state. For smaller values of $k_s$ that satisfy \eqref{eq.tertiaryControlGainCondition} the frequency response was poor, however, we observed that there was improved power sharing between the inverters. As observed through the simulations and from Remark~\ref{rmk1}, there is an inherent trade-off between the gains $k_s$ and $k_t$. The ratio of the gains $(k_s, k_t)$ can be tuned to achieve the desired objectives.

\section{Conclusions} \label{sec.Conclusion}
This paper introduced a decentralized secondary control scheme that is robust to clock synchronization errors and is able to restore the frequency in inverter-based islanded microgrids. A unified control-theoretic approach was utilized to systematically show that the microgrid dynamics are stable and reach a steady-state. The proposed DSC scheme is agnostic to the inverter-type and can be used seamlessly with GFM or GFL inverter architectures. Criteria for designing the secondary control gains such that they provide good performance for frequency regulation while ensuring fair power sharing are also discussed. Current research efforts are focused on testing the novel DSC scheme under mixed setups of GFM and GFL inverters as well as validating the results in our microgrid hardware testbed comprised of several inverter-based resources.
	\bibliographystyle{IEEEtran}
	\bibliography{myabrv,references}
	
\end{document}